\documentclass[prl,twocolumn,groupedaddress]{revtex4-1}

\usepackage{epsfig}
\usepackage{graphicx}
\usepackage{color}

\usepackage[caption=false]{subfig}

\begin{document}
\renewcommand{\thefigure}{\arabic{figure}}
\setcounter{figure}{0}

\bibliographystyle{apsrev}

\title{Mapping the nano-Hertz gravitational wave sky}

\author{ \surname{Neil} J. Cornish}
\affiliation{Department of Physics, Montana State University, Bozeman,
MT 59717}
\author{ \surname{Rutger} van Haasteren}
\affiliation{Jet Propulsion Laboratory, California Institute of Technology, 4800 Oak Grove Drive, Pasadena, CA 91106, USA }

\begin{abstract}
We describe a new method for extracting gravitational wave signals from pulsar timing data. We show that any
gravitational wave signal can be decomposed into an orthogonal set of sky maps, with the number of maps equal to
the number of pulsars in the timing array. These maps may be used as a basis to construct gravitational wave templates
for any type of source, including collections of point sources. A variant of the standard Hellings-Downs correlation
analysis is recovered for statistically isotropic signals. The template based approach allows us to probe
potential anisotropies in the signal and produce maps of the gravitational wave sky.
%, and in addition, offers significant computational
%savings over a correlation based analysis as there are no large correlation matrices to invert.
\end{abstract}

\maketitle

Millisecond pulsars emit pulse trains with a timing stability that rivals the best atomic clocks. After taking into account
relative motion and propagation effects with an accurate timing model, the best timed pulsars have timing residuals
of tens of nanoseconds. Gravitational waves will impart a distinct variation in the timing residuals~\cite{Estabrook:1975, Sazhin:1978, Detweiler:1979wn}
that allows us to separate this signal from noise~\cite{Hellings:1983fr, Foster:1990}. Here we present a radically different approach
for analyzing timing data from an array of pulsars based on sky maps that allows us to detect anisotropy, and offers new insight into the
geometrical underpinnings of the analysis.

The likelihood of observing timing residuals ${\delta {\bf t}}$ in the presence
of a gravitational wave signal ${\bf h}$, given a timing model with parameters $\vec{\xi}$, and a noise model with parameters $\vec{\phi}$
is given by~\cite{vanHaasteren:2012hj}
\begin{eqnarray}\label{like}
&& p(\delta{\bf t}\vert {\bf h}, \vec{\xi}, \vec{\phi}) =  \frac{1}{\sqrt{(2\pi)^n {\rm det}{\bf C}}} \times \nonumber \\
&& \exp\left(-\frac{1}{2} (\delta{\bf t} - {\bf F}{\bf h}-{\bf M} \vec{\xi})^T {\bf C}^{-1} (\delta{\bf t} - {\bf F}{\bf h}-{\bf M} \vec{\xi})\right),
\end{eqnarray}
where ${\bf F}$ is the network response operator, ${\bf M}$ is the design matrix for the timing model, and ${\bf C}$ is the noise covariance matrix. Here $n$ is
the total number of data points. It is convenient to introduce an upper triangular Cholesky decomposition for the noise: ${\bf C}^{-1}  = {\bf Q}^T {\bf Q}$, and whiten the data by replacing
$\delta{\bf t} \rightarrow  {\bf Q} \delta{\bf t}$, ${\bf F} \rightarrow {\bf Q} {\bf F}$ and ${\bf M} \rightarrow {\bf Q} {\bf M}$. To avoid introducing
additional notation, we will simply refer to ${\bf Q} \delta{\bf t}$ as $\delta{\bf t}$ {\it etc} in what follows.

The network response ${\bf s} = {\bf F}{\bf h}$ can be cast in matrix form by introducing a pixelization of the sky with $N$ equal-area pixels, such that the gravitational
wave signal in the $\hat{n}$ direction, $(h^+(\hat{n}), h^\times(\hat{n})$ can be represented by the element $h_n = (h^+_n, h^\times_n)$ of the column
vector ${\bf h}$. The gravitational wave induced timing residuals in the $j^{\rm th}$ pulsar in array of $N_p$ pulsars can then be written as~\cite{Cornish:2013nma}
\begin{equation}\label{resp}
s_j = \sum_{n=1}^N  F^+_{j n} h^+_n + F^\times_{j n} h^\times_n \, .
\end{equation}
In matrix form ${\bf s} = {\bf F}{\bf h}$, ${\bf s}$ is a $N_p \times 1$ column vector,  ${\bf F}$ is the $N_p \times 2N$ network response matrix, and ${\bf h}$ is a $2N \times 1$ column vector.
%Since we ultimately want to take the limit $N \rightarrow \infty$, it will prove convenient to divided each element of ${\bf F}$ by $\sqrt{N}$, and multiply each
%element of ${\bf h}$ by $\sqrt{N}$.
There is a separate copy of $s_j$ for each time sample, but we suppress this additional index in an effort to keep the notation compact.

The geometrical properties of the network response operator are revealed by performing a singular value decomposition ${\bf F} = {\bf U} {\bf \Sigma} {\bf V}^T$.
The $2N$ columns of ${\bf V}$ with non-zero singular values are the sky basis vectors ${\bf v}_{(k)}$, and the $N_p$ rows of ${\bf U}$ with non-zero singular values
are the range vectors $\bf{u}_{(k)}$.  Here we are using notation where indices in parentheses label which vector we are referencing,
while indices without parentheses label the components of the vectors. The sky basis vectors and range vectors are related via ${\bf F}  {\bf v}_{(k)} = \sigma_k  { \bf u}_{(k)},$
%\begin{equation}
%{\bf F}  {\bf v}_{(k)} = \sigma_k  { \bf u}_{(k)}\,,
%\end{equation}
where $\sigma_k$ is the singular value for the $k^{\rm th}$ component of the decomposition. One convenient method for computing these quantities is to use the Healpix sky
pixelization~\cite{Gorski:2004by}. For moderate sized arrays of pulsars (say tens to hundreds), the solution for the singular values and range vectors rapidly converges to a unique solution
as the number of pixels is increased. We found that a Healpix $N_{\rm side}=32$, which has $N=12288$ equal-area pixels, was sufficient. Note that the sky pixelization is
merely a computational convenience. Any other basis can be used to represent the network response operator and the same range vectors will result. This has
been confirmed by Gair {\it et al}~\cite{Gair:2014rwa} using a spherical harmonic basis.

\begin{figure}[ht]
\includegraphics[scale=0.5]{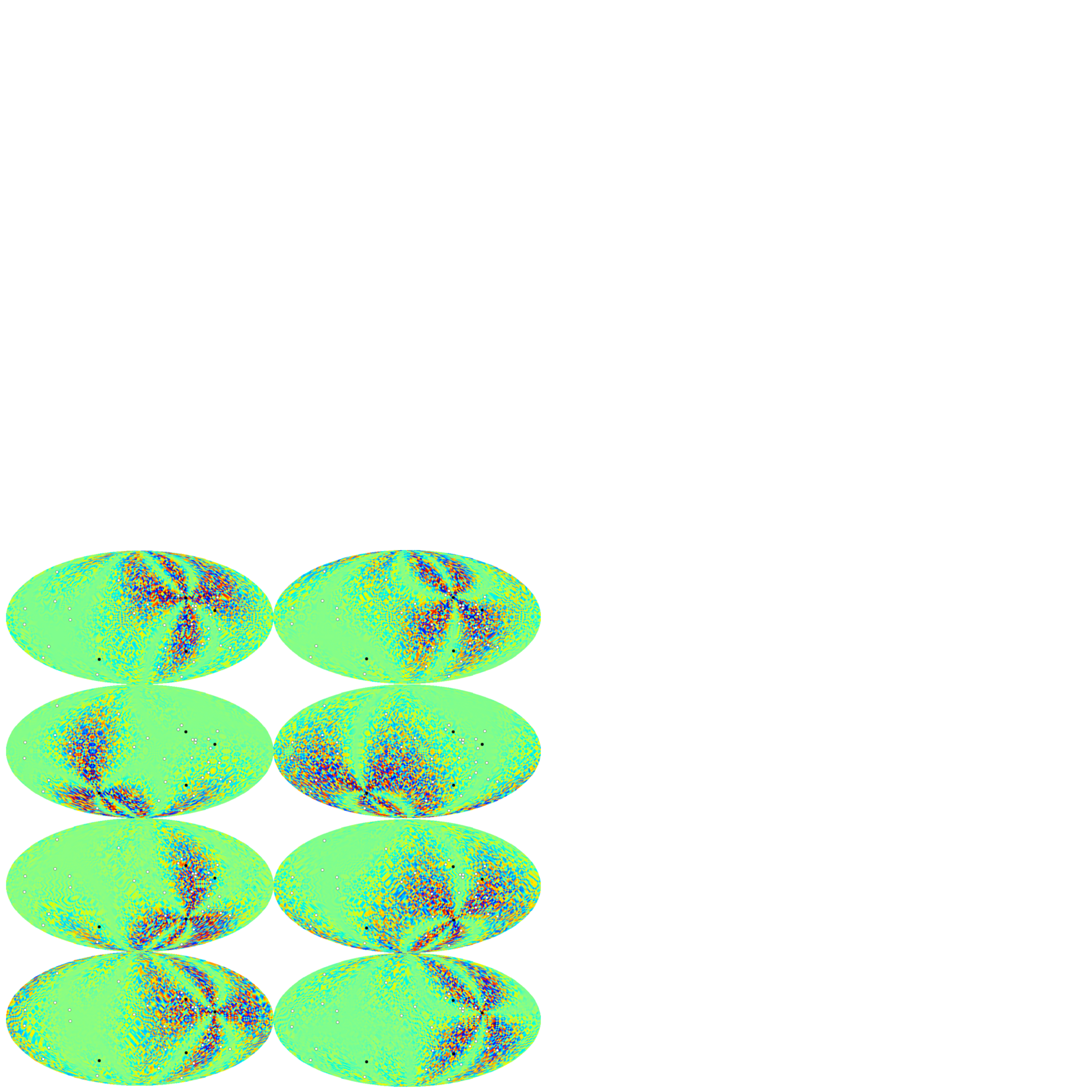}
\caption{The four dominant Pulsar term sky basis vectors for the IPTA, shown in descending order of their singular values, with the plus polarization on the left and the cross
polarization on the right. Here the signal was assume to be monochromatic, with a frequency of $f=10^{-8}$ Hz.
The location of the 36 pulsars in the array are show as black or white dots. The black dots indicate the locations of the best timed pulsars, J1939+2134,
J0437-4715, J1713+0747 and J1909-3744. As expected, the RMS sky map amplitudes are simply the antenna patterns for each of these pulsars.}
\label{fig:pbasis}
\end{figure}

Note that any gravitational sky ${\bf h}$ can be decomposed into a portion that registers a response in the pulsar timing array, ${\bf h}_{\rm obs} =  \gamma_k {\bf v}_{(k)}$, with
$\gamma_k = {\bf v}_{(k)}\cdot {\bf h}$,  and a portion that lies in the null space of the response operator, ${\bf h}_{\rm null} = {\bf h} - {\bf h}_{\rm obs}$. We have ${\bf s} = {\bf F}{\bf h}
={\bf F}{\bf h}_{\rm obs}= \gamma_k \sigma_k  { \bf u}_{(k)}$. The $N_p$ amplitudes $\gamma_k$ of the sky basis vectors provide a natural set of variables with which to model any
gravitational wave signal. One significant complication is that the network response operator has two terms - the Earth term and the pulsar term ${\bf F}   = {\bf F}^{\rm E} + {\bf F}^{\rm P}.$
%\begin{equation}
%{\bf F}   = {\bf F}^{\rm E} + {\bf F}^{\rm P} \, .
%\end{equation}
The Earth term is a simple matrix with components that vary slowly across the sky, while the pulsar term is a time delay operator that acts on the phase of the gravitational
wave, imparting time delays in the phase seen by the $k^{\rm th}$ pulsar of the form $\Phi(t - L_k(1-\cos\mu_k))$, where $L_k$ is the distance to the pulsar, and $\mu_k$ is the
angle between the line of sight to the pulsar and the gravitational wave signal. For monochromatic sources this leads to a frequency dependent phase shift
at each point on the sky. For evolving sources the behavior is more complicated. In either case, the components of ${\bf F}^{\rm P}$ oscillate rapidly across the sky, with a
coherence length that is typically much less than one degree. Since the two components of the network response matrix behave so differently, it makes sense to decompose
each of them into their own set of sky basis functions and range vectors (while ${\bf F}^{\rm E}$ and ${\bf F}^{\rm P}$ share the same range, they have different sets of
orthonormal range vectors). The range vectors for the pulsar term have components ${u}^{\rm P}_{(k)i} = \delta_{ki}$ - in other words, the pulsar term produces an uncorrelated
response in the array. The corresponding components of the sky basis vectors for the pulsar term have root-mean-square amplitudes equal to the antenna patterns for that pulsar:
$({v}^{\rm P}_{(k)n})_{\rm RMS} = (F^+_{kn}, F^\times_{kn})$. If all the pulsars in the array are equally sensitive, then the singular values for the pulsar term $\sigma^P_k$ are
all equal. More generally, after applying the Cholesky whitening, the singular values scale inversely with the noise level in each pulsar: $\sigma^P_k \sim S_{k}(f)^{-1/2}$. 
The four dominant pulsar-term sky basis vectors for the International Pulsar Timing Array (as used in the first IPTA mock data challenge~\cite{IPTA}), broken out into the two gravitational
wave polarization states ${ \bf v}^{\rm P}_{(k)} = ({ \bf v}^{\rm P+}_{(k)}, { \bf v}^{\rm P\times}_{(k)})$, are shown in Figure \ref{fig:pbasis}.

\begin{figure}[ht]
\includegraphics[scale=0.5]{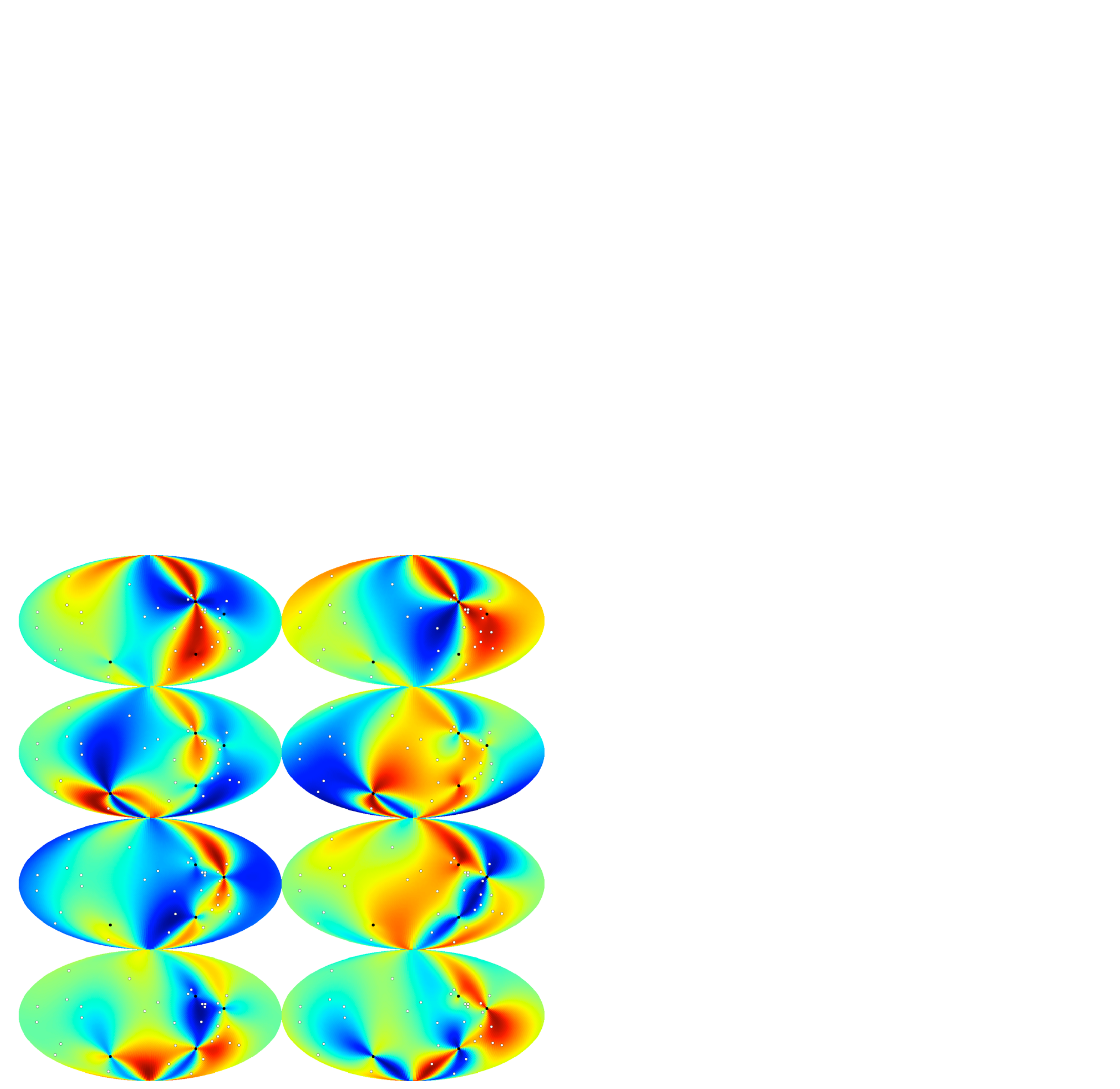}
\caption{ The four dominant Earth term sky basis vectors for the IPTA, shown in descending order of their singular values, with the plus polarization on the left and the cross
polarization on the right.}
%The location of the 36 pulsars in the array are show as black or white dots. The black dots indicate the locations of the best timed pulsars, J1939+2134,
%J0437-4715, J1713+0747 and J1909-3744.}
\label{fig:ebasis}
\end{figure}

%The range and sky vectors for the Earth-term behave very differently than the pulsar term.
The Earth-term range vectors include correlations between pulsars, and the singular values vary significantly, even
when all the pulsars are equally sensitive (the singular values for an equal sensitivity array typically vary by factors of 10 to 100 from largest to smallest, depending on the specific
geometry of the array).  The four dominant Earth term sky basis vectors for the International Pulsar Timing Array are shown in Figure \ref{fig:ebasis}. The four best timed pulsars dominate these
maps, the shapes of which can be understood from the observation that the sky basis for a single pulsar is proportional to its antenna pattern~\cite{joe}.

In a Bayesian analysis, after specifying priors for the model parameters $\gamma_k, \vec{\xi}, \vec{\phi}$, numerical techniques such as Markov Chain Monte Carlo can be
used to estimate the posterior distribution function for the sky map. Before discussing such an analysis, it is instructive to look at the maximum likelihood (ML) solution under the assumption that the noise model
and timing model is known (so $\vec{\xi}=0$). If the distance to each pulsar was known to exquisite accuracy, it would be possible to reconstruct both the combined Earth-term map
and pulsar-term map. With enough sensitivity over a range of frequencies it may even be possible to recover the distances to the pulsars from the response to the GW signal~\cite{Corbin:2010kt}.
If the distances to the pulsars are known, the timing residual
\begin{equation}\label{dt0}
\delta{\bf t} = \delta{\bf t}^{\rm GW}  + \delta{\bf t}^{\rm n} = {\bf F}{\bf h} + {\bf n}
\end{equation}
yields the ML solution
\begin{equation}\label{MLF}
{\bf h}^{\rm ML} =  {\bf h}_{\rm obs} + {\bf F}^\# {\bf n} \, ,
\end{equation}
where ${\bf F}^\#={\bf V} {\bf \Sigma}^{\#} {\bf U}^T$ is the pseudo-inverse of ${\bf F}$, and ${\bf \Sigma}^{\#}$ is the
pseudo inverse of ${\bf \Sigma}$, which is found by replacing the non-zero diagonal elements of ${\bf \Sigma}$ by their reciprocal values.
The ML solution for the sky basis amplitudes is given by $\gamma^{\rm ML}_k = {\bf v}_{(k)}\cdot {\bf h}^{\rm ML}$.
For timing residuals dominated by zero mean Gaussian noise, the pixels in the reconstructed sky maps follow a multi-variate Gaussian distribution with
\begin{eqnarray}\label{skynoise}
{\rm E}(h_i^{\rm ML}) &=& 0 \nonumber \\
{\rm E}(h_i^{\rm ML} h_j^{\rm ML}) &=& \sum_k \frac{ v_{(k)i} v_{(k)j}}{(\sigma_k)^2}\, ,
\end{eqnarray}
or equivalently,
\begin{eqnarray}\label{stocn}
{\rm E}(\gamma^{\rm ML}_k) &=& 0 \nonumber \\
{\rm E}(\gamma^{\rm ML}_k\gamma^{\rm ML}_l) &=&\frac{1}{\sigma_k^2}\,  \delta_{kl}\, ,
\end{eqnarray}
with no summation on $k$ in the last expression. 
The noise in the reconstruction is dominated by sky maps with small singular values. In a Bayesian analysis this problem can be avoided by using a trans-dimensional MCMC to select the sub-set of the sky basis vectors that
optimally balances model fidelity against model complexity. In a frequentist analysis we can achieve a similar result by using a
low-rank approximation to the pseudo-inverse, ${\bf F}^\#$, which is found by replacing the largest diagonal elements of ${\bf \Sigma}^\#$ by zero.
The reconstruction can be further improved by specifying suitable priors on the sky basis amplitudes.\\

\begin{figure}[ht]
\centerline{\includegraphics[scale=0.5]{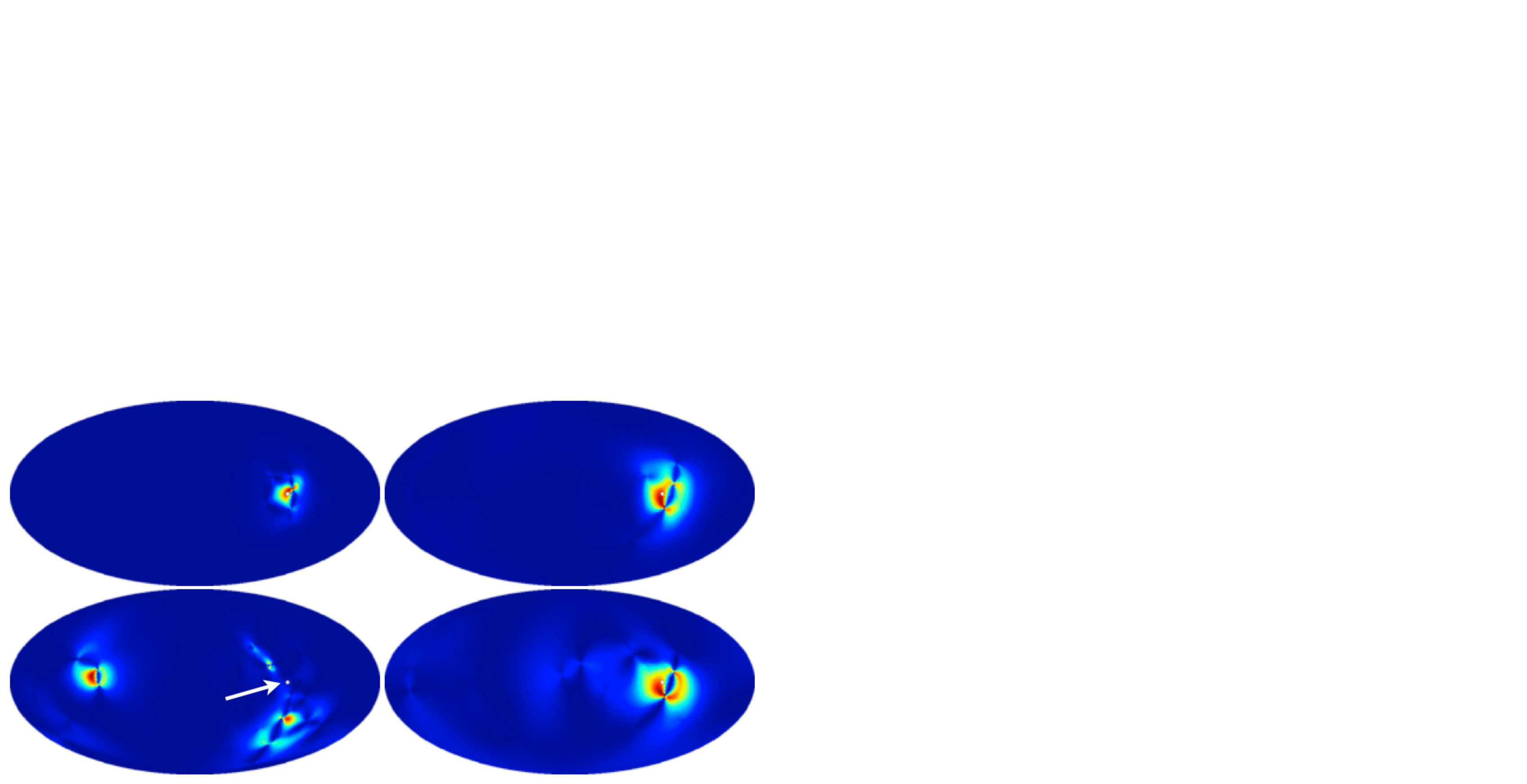}}
\caption{Sky map reconstruction of $h_+^2+h_\times^2$ for a point source. The first column uses the full 36 Earth-term sky basis vectors for the IPTA, while the second column uses
the 10 basis functions with the largest singular values. The first row is for the Earth term contribution, while the second row includes pulsar and noise
contributions. The whitened signal power was set equal to the whitened noise level. The white circles show the location of the point source, which is indicated
by an arrow in the bottom left panel.}
\label{fig:point}
\end{figure}

In the event that the pulsar distances can not be determined to sufficient accuracy, we can attempt to recover the Earth-term sky.
To do this, we first break the timing residual out into the contribution from the Earth-term, pulsar-term, and timing noise:
\begin{equation}\label{dt}
\delta{\bf t} = \delta{\bf t}^{\rm E} + \delta{\bf t}^{\rm P}  + \delta{\bf t}^{\rm n} = {\bf F}^{\rm E} {\bf h} + {\bf F}^{\rm P} {\bf h} + {\bf n} \, .
\end{equation}
In this case we treat the pulsar term as an additional noise source in the ML reconstruction. The Earth-term
ML solution is given by
\begin{equation}\label{MLE}
{\bf h}^{\rm E,ML} =  {\bf h}^{\rm E}_{\rm obs} + {\bf F}^{\rm E\#} {\bf F}^P {\bf h} + {\bf F}^{\rm E\#} {\bf n} \, ,
\end{equation}
where ${\bf F}^{E\#}={\bf V}^{\rm E} {\bf \Sigma}^{{\rm E}\#} {\bf U^{\rm E}}^T$ is the pseudo-inverse of ${\bf F}^E$.
The ML solution for the Earth term sky basis amplitudes is given by $\gamma^{\rm ML}_k = {\bf v}^{\rm E}_{(k)}\cdot {\bf h}^{\rm ML}$.
The noise in the reconstruction $\delta {\bf h} = {\bf h}^{\rm E,ML} - {\bf h}^{\rm E}_{\rm obs}$ has contributions from the
instrument noise and the pulsar term. Figure 3 shows full and reduced rank reconstructions of a point source. The full rank
reconstruction with pulsar term and noise is badly corrupted, while the reduced rank reconstruction is not.

{\em Statistically Isotropic Signals} Since the sky template analysis is entirely general, it can be used in place of the standard cross-correlation analysis for
isotropic stochastic signals. With a suitable parameterized prior on the amplitudes $\gamma_k$, defined below, the model shares the same dimensionality
as the correlation analysis. However, the template based analysis offers the distinct computational advantage of avoiding the inversion of large correlation
matrices when computing the likelihood.

A statistically isotropic stochastic signal is fully characterized by the expectation values for the sky-pixel amplitudes:
\begin{eqnarray}\label{skystoc}
{\rm E}(h_i) &=& 0 \nonumber \\
{\rm E}(h_i h_j) &=& \frac{1}{2} S_h \delta_{ij} \, .
\end{eqnarray}
The factor of one-half comes from averaging over the polarization angle.
If the pulsars distances were known, we could work with the sky-basis vectors for the full response matrix and write
\begin{equation}
\delta {\bf t} = \sigma_k ({\bf v}_{(k)}\cdot {\bf h}) {\bf u}_{(k)}+{\bf n} \, ,
\end{equation}
from which it then follows that the timing residuals would be described by a multi-variate Gaussian distribution with
\begin{eqnarray}\label{tstoc0}
{\rm E}(\delta t_i) &=& 0 \nonumber \\
{\rm E}(\delta t_i \delta t_j) &=& \frac{1}{2} S_h \sigma_k^2 u_{(k)i} u_{(k)j} +  \delta_{ij} \, .
\end{eqnarray}
The expression for the cross correlation can be put in a more familiar form if we recall that what we really have in (\ref{tstoc0}) is ${\rm E}(({\bf Q} \delta {\bf t})_i ({\bf Q} \delta {\bf t})_j)$.
In the frequency domain, under the assumption that the noise in each pulsar is uncorrelated, the noise correlation matrix is diagonal: $C_{ij}(f) = S_i(f) \delta_{ij}$, where $S_i(f)$ is the noise in the $i^{\rm th}$ pulsar,
and $Q_{ij}(f) = S_i^{-1/2}(f) \delta_{ij}$. Undoing the Cholesky whitening we find
\begin{equation}
{\rm E}(\delta t_i \delta t_j)_{\rm Colored} =  S_h(f) \beta_{ij}+  S_i(f)  \delta_{ij} \, ,
\end{equation}
where the correlation matrix
\begin{equation}
\beta_{ij} =\frac{ \sigma_k^2(f)}{2} {u_{(k)i} u_{(k)j}}{(S_i(f) S_j(f))^{1/2}}
\end{equation}
is closely related to the Hellings-Downs (H\&D) correlation matrix~\cite{Hellings:1983fr}. It differs since here we are considering the ideal case where both the pulsar-term and the
Earth-term can be treated coherently. Using (\ref{MLF}), it can be shown that the expectation values for the amplitudes of the ML sky basis vectors are given by
\begin{eqnarray}\label{sfull}
{\rm E}(\gamma^{\rm ML}_k) &=& 0 \nonumber \\
{\rm E}(\gamma^{\rm ML}_k\gamma^{\rm ML}_l) &=& \left( \frac{S_h}{2} + \frac{1}{\sigma_k^2} \right) \delta_{kl}    \, .
\end{eqnarray}
Equations (\ref{tstoc0}) and (\ref{sfull}) contain the same information but package it differently. In terms of the timing correlations in (\ref{tstoc0}), the gravitational wave signal
and the instrument noise can be separated as they have different correlation matrices - the gravitational wave signal is correlated between pairs of pulsars while the noise
is not. In terms of the amplitude correlations in (\ref{sfull}) the gravitational wave signal and the instrument noise can be separated as they enter the different sky maps
with different strengths. 

More realistically, when the pulsar distances are not known to high accuracy we have to split the response into Earth-term, pulsar-term and noise contributions
%\begin{eqnarray}
%\delta {\bf t}^{\rm E} &=& \sigma_k^{\rm E} ({\bf v}^{\rm E}_{(k)}\cdot {\bf h}) {\bf u}_{(k)}^{\rm E}  \nonumber \\
%\delta {\bf t}^{\rm P} &=& \sigma_k^{\rm P} ({\bf v}^{\rm P}_{(k)}\cdot {\bf h}) {\bf u}_{(k)}^{\rm P}  \nonumber \\
%\delta {\bf t}^{\rm n} &=& {\bf n}\, ,
%\end{eqnarray}
which leads to a multi-variate Gaussian distribution for the timing residuals:
\begin{eqnarray}\label{tstoc}
{\rm E}(\delta t_i) &=& 0 \nonumber \\
{\rm E}(\delta t_i \delta t_j) &=& \frac{S_h}{(S_i S_j)^{1/2}} \alpha_{ij} +  \delta_{ij}\, ,
\end{eqnarray}
where $\alpha_{ij}$ is the H\&D~\cite{Hellings:1983fr} correlation matrix
\begin{eqnarray}
\alpha_{ij} &=&\frac{1}{2}\left( (\sigma_k^{\rm E}(f))^2 u^{\rm E}_{(k)i} u^{\rm E}_{(k)j}(S_i(f) S_j(f))^{1/2} + \delta_{ij} \right) \nonumber \\
&=& \frac{1}{2} + \frac{3}{2} \kappa_{ij} \ln \kappa_{ij} -\frac{1}{3} \kappa_{ij} + \frac{1}{2} \delta_{ij} 
\end{eqnarray}
with $\kappa_{ij} = (1-\cos(\theta_{ij}))/2$, where $\theta_{ij}$ is the angle between the line of sight to pulsars $i,j$.
Note that the cross term ${\rm E}(\delta t^{\rm E}_i \delta t^{\rm P}_j)$ vanishes since ${\bf v}^{\rm E}_{(k)}\cdot {\bf v}^{\rm P}_{(l)} = 0$, which is a consequence of
the pulsar-term sky maps oscillating rapidly across the sky and integrating to zero against the Earth-term sky maps which are smooth functions.
Undoing the Cholesky whitening and working in the frequency domain we get
\begin{equation}
{\rm E}(\delta t_i \delta t_j)_{\rm Colored} = S_h(f) \alpha_{ij}+  S_i(f)  \delta_{ij} \, ,
\end{equation}
It is interesting to note that the scaled range vectors $\bar{u}^{\rm E}_{(k)(i)} = u^{\rm E}_{(k)(i)}/(\sigma^{\rm E}_k S_i^{1/2})$ diagonalize the Earth-term of the H\&D correlation matrix.
The fact that the range vectors, which can be used to describe any GW signal, form the H\&D correlation matrix explains why this quantity,
which was originally derived for isotropic skies, is also relevant to point sources~\cite{Cornish:2013aba}.

%This provides an alternative way to derive the range vectors, however a pixel basis or spherical
%harmonic basis are still needed to find the sky-basis maps.

%The expectation values for the correlations in the Earth-term, pulsar-term and noise contributions to the ML Earth-term sky basis amplitudes for an isotropic
%stochastic signal are given by
%\begin{eqnarray}\label{ge}
%{\rm E}(\gamma^{\rm E}_k \gamma^{\rm P}_l) &=& 0 \nonumber \\
%{\rm E}(\gamma^{\rm E}_k \gamma^{\rm n}_l) &=& 0 \nonumber \\
%{\rm E}(\gamma^{\rm P}_k \gamma^{\rm n}_l) &=& 0 \nonumber \\
%{\rm E}(\gamma^{\rm E}_k \gamma^{\rm E}_l) &=& \frac{S_h}{2} \delta_{kl} \nonumber \\
%{\rm E}(\gamma^{\rm n}_k \gamma^{\rm n}_l) &=& \frac{1}{(\sigma^{\rm E}_k)^2} \delta_{kl} \nonumber \\
%{\rm E}(\gamma^{\rm P}_k \gamma^{\rm P}_l) &=& S_h \frac{ (\sigma_q^{\rm P})^2 u^{\rm E}_{(k)q} u^{\rm E}_{(l)q}}{2 \sigma^{\rm E}_k \sigma^{\rm E}_l} 
%= \frac{S_h}{2} \delta_{kl} \, .
%\end{eqnarray}
%Combining these contributions we have

The amplitudes of the sky basis maps for the Earth-term and pulsar-term each have the correlation structure
\begin{eqnarray}\label{searth}
{\rm E}(\gamma_k) &=& 0 \nonumber \\
{\rm E}(\gamma_k\gamma_l) &=& \frac{S_h}{2} \delta_{kl}    \, .
\end{eqnarray}
The Earth-pulsar cross terms vanish.
In a Bayesian analysis the correlation structure (\ref{searth}) serves as a prior on the $\gamma_k$.  Analytically marginalizing over the pulsar-term contribution
adds a diagonal component to the noise correlation matrix proportional to $S_h/2$. Analytically marginalizing over the Earth-term contribution results in the standard
H\&D correlation analysis. Alternatively, we can numerically marginalize over the Earth-term GW templates $\delta {\bf t}^{\rm GW} = \gamma^{\rm E}_k \sigma_k {\bf u}^{\rm E}_{(k)}$,
thereby avoiding the costly step of inverting a large correlation matrix when computing the likelihood. The templates can be generated in the
Fourier domain then transformed to the time domain and interpolated to match the un-even sampling of the data~\cite{Lentati:2012xb}. Updates to the noise
model parameters $\vec{\phi}$ introduce a minor complication as they alter the Cholesky whitening, which changes the sky basis vectors, singular values and range vectors.
Since the updated Earth-term response matrix shares the same null space, column space and range as the original response matrix the new sky basis
vectors and range vectors can be expressed as linear combinations of the original vectors, and the amplitudes of the sky basis amplitudes can be mapped
to the new basis.
%Rather than work with an ever changing basis, it is simpler to adopt a reference basis corresponding to a particular instance of the
%noise model and to account for the changing noise correlation matrix in the calculation of the likelihood (which will pick up off-diagonal terms when
%the noise model differs from the reference model).
The model dimension for the template based analysis matches that of the standard Hellings-Downs
cross-correlation analysis, but could offer dramatic computational savings since the only matrices that need to be factorized or inverted are block diagonal.

{\em Anisotropic Signals} The sky template approach is ideally suited to studying anisotropic signals. Equation (\ref{MLE}) provides a general maximum
likelihood reconstruction of any signal, but this can be improved upon in a Bayesian analysis that builds in priors on the sky-basis amplitudes. The key
signature of an anisotropic signal is a non-diagonal (though diagonal dominant) correlation matrix $\gamma_k \gamma_l$. A variant of the
isotropic search described above, but with a weaker prior on the correlation matrix, can be used to detect anisotropies in the nanoHz gravitational
wave sky.

\section*{Acknowledgments}
We would like to thank Joe Romano for several informative discussions. We thank Jonathan Gair, Stephen Taylor,  Joe Romano and Chiara Mingarelli for sharing 
a draft of their work using spherical harmonics to characterize anisotropic gravitational signals with pulsar timing arrays, and for verifying several of our key
results using their formalism. NJC appreciates the support of NSF grant
PHY-1306702. RvH is supported by NASA Einstein Fellowship grant PF3-140116.
This work was partially carried out at the Jet Propulsion Laboratory, California
Institute of Technology, under contract to the National Aeronautics and Space
Administration. Copyright 2014.

\end{document}